# Enhanced fault-tolerance in biomimetic hierarchical materials – a simulation study

Seyyed Ahmad Hosseini*, Paolo Moretti, Michael Zaiser

*Institute of Materials Simulation (WW8), Friedrich-Alexander-Universität Erlangen-Nürnberg (FAU), Dr.-Mack-Str. 77, 90762 Fürth, Germany*

**Abstract**

Hierarchical microstructures are often invoked to explain the high resilience and fracture toughness of biological materials such as bone and nacre. Biomimetic material models inspired by those structural arrangements face the obvious challenge of capturing their inherent multi-scale complexity, both in experiments and in simulations. To study the influence of hierarchical microstructural patterns in fracture behavior, we propose a large scale three-dimensional hierarchical beam-element simulation framework, where we generalize the constitutive behavior of Timoshenko beam elasticity and Maximum Distortion Energy Theory failure criteria to the complex case of hierarchical networks of approximately 5 million elements. We perform a statistical study of stress-strain relationships and fracture surface mophologies, and conclude that hierarchical systems are capable of arresting crack propagation, an ability that reduces their sensitivity to pre-existing damage and enhances their fault tolerance compared to reference generic fibrous materials.

## 1. Introduction

Hierarchically structured materials display a self-similar organization, in which comparable microstructural features are reproduced on multiple scales. A global tendency of biological systems to be organized in a hierarchical modular fashion is evident, for instance, in the case of collagen, which contains hierarchically structured patterns, from the moleculecular scale of amino-acids chains, through microfibrils and fibers, up to hierarchical fiber bundles. This arrangement ensures enhanced fracture over assemblies of isolated collagen molecules [1], and is believed to play a major role, for instance, in bone fracture [2, 3].

---

*Corresponding author
 *Email address:* ahmad.hosseini@fau.de (Seyyed Ahmad Hosseini)

Similar considerations have been proposed to explain and characterize the cellular structure of wood [4], and hierarchical lamellar microstructures of mollusc shells (nacre)[5]. In this broad context, hierarchical structures are believed to control load distribution and contain damage, thus enhancing resilience even in materials with brittle constituents [6, 7].

The possibility of engineering hierarchical biomimetic structures, which are fault tolerant regardless of the inherent unreliability of their constituents, or the fluctuations in their constitutive behavior, is of course very appealing from the point of view of biomaterial synthesis and design. While the preponderant role of the hierarchical structure is a gross simplification when addressing real biological materials like bone, it could prove useful in the context of biomaterials, where one wants to control fracture behavior, rather than modelling a complex biological system. The mechanisms of synthesis and growth of materials which are known to exhibit hierarchical microstructures, such as nacre, can sometimes be observed in controlled lab environments [8]. However, experimental growth often is a nanoassembly process, which one cannot expect to reconstruct a full hierarchical organization, which extends well beyond the microscale. An alternative route to synthesize and characterize hierarchical structures involves numerical modeling. Simulation methods allow for the modeling and tuning of mechanically loaded structures in view of applications, which may for instance involve techniques of additive manufacturing that operate at scales much larger than particle self assembly.

Here we propose a large scale numerical study of bulk fracture in 3D fibrous hierarchical structures and show that even under minimal assumptions systems of this type exhibit a unique fracture behavior, where the nucleation of micro-cracks is not followed by their expansion or propagation. While previous studies have highlighted this behavior in simplified 2D models [9] and/or resorting to simplified constitutive laws [10, 11, 12], our work here is the first where a realistic, large-scale, beam-element simulation framework is implemented, and the first to provide exhaustive evidence of the enhanced fault tolerance of these structures.

Regarding numerical models of failure in hierarchical systems, fiber bundle models are among the simplest tools to examine collective phenomena in deformation and fracture [13] and have been widely used to study disordered fibrous materials. The material is modelled as a collection of load-bearing fibers that, in the simplest example of a brittle fiber bundle, elastically deform until they fail at a critical load. The local constitutive behavior can be modified, beyond the ideal brittleness of individual fibers, to incorporate gradual damage accumulation [14], creep [15], and plasticity [16]. Importantly, hierarchical generalizations of such models have also been proposed [17, 18]. However, fiber bundle models are not intended to investigate how stress is re-distributed across a material, nor can they clearly tell if failure is caused by the growth of a local nucleus triggered by stress concentrations, damage percolation across the entire system, or some intermediate mechanism. As a consequence, within fiber bundle models have devoted comparatively little attention to the fundamental topic of the nature of the failure process [13].

On the lowest structural level, fracture of hierarchical materials can be modeled using atomistic methods. These simulations may subsequently be used to parametrize mesoscale models that describe behavior at higher hierarchical levels [19]. Alternatively, they may be taken to include several hierarchical levels in extremely large-scale simulations [1], where computational cost can limit the access to larger sizes.

Lattice/network models represent a viable alternative to efficiently capture multiple hierarchical levels and length scales. Discrete lattice models for materials fracture were introduced to study the statistical effects of microstructural disorder (more precisely, fluctuating local strengths) [20]. In models like the random-fuse model (RFM), the random-spring model and the random-beam model, materials are represented as networks of discrete elements that transmit scalar, vector, or tensor loads. Such models serve as paradigms for fracture as a multiscale process, capturing its key features such as the interplay between local failure processes, microstructural heterogeneity (represented, e.g., in terms of locally random network architecture or locally random failure thresholds of links), and system-spanning interactions as the load re-distributes across the network. Such models attain a level of simplicity that allows for large-scale simulations and statistically meaningful predictions [21]. Original lattice models of biometic structures, inspired for instance by nacre, resorted to RFM simulations and their simplifying assumptions of scalar loads and scalar constitutive laws [22], though without considering the material's hierarchical structure [23]. These models were later generalized to the case of tensor-load descriptions, where both elastic equilibrium equations and constitutive laws take into account linear and angular momenta [24]. More recently similar beam lattice models have been used to explore optimal network arrangements in view of fracture behavior [25] and in particular to investigate biomimetic 2D hierarchical structures.

In this study, we introduce a 3D hierarchical beam lattice model to explore how the hierarchical organization affects the fracture properties of biomimetic fibrous materials. We compare the crack sensitivity of hierarchical structures to that of reference, non-hierarchical systems. Finally, we compare the characteristics of failure and morphology of fracture surfaces to understand how the microstructure contributes to the higher mechanical performance of these systems.

## 2. Methods

Our lattice models is based on a cubic lattice of interconnected beams clamped together at their intersections. The points where beams are mutually connected are called nodes, and the number of nodes at each lattice edge is referred to as the lattice size $L$. The simulations in this study are performed on 3D lattice systems of size $L = 128$, and tensile loads directed along the $z$ axis (see Fig. 1). Beams with the same orientation as the load are called load-carrying (LC) beams, their number is denoted as $N_{\mathrm{LC}}$, and a set of $l$ connected LC beams is referred to as a LC fiber of length $l$. In contrast, beams oriented perpendicular to the load are called cross-link (CL) beams, their number is denoted as $N_{\text{CL}}$, and a set of $l$ connected CL beams is referred to as a CL *connector* of length $l$. Finally, a surface with normal on the $xy$ plane, where all CL beams are missing is denoted as a *gap*. According to our nomenclature, connectors are extended chains that contribute to load redistribution, while gaps are extended surfaces that interrupt it.

*2.1. Construction of 3D hierarchical and non hierarchical beam lattices*

In order to visualize the structure of a deterministic hierarchical beam lattice (DHBL) we devise an iterative "bottom-up" method, as shown in Fig. 1. A level-1 module, or generator, is the basic building block of the hierarchical structure, constituted by a CL plane and 8 LC beams. At every iterative step $n$, this structural pattern is replicated, in the form of a (periodically continued) CL plane of linear size $2^n$ and 8 LC super-beams, each corresponding to a level-$(n-1)$ system.

This construction results in a hierarchical arrangement of modules separated by gaps and connected by CL connectors. In the DHBL pattern, as well as in analogous 2D structures [10] the linear sizes of gaps are power-law distributed. The structure of the model is deterministic in the sense that its connectivity features at every location can be deduced from the location itself, with no random deviations. Stochastically shuffled versions of these structures have been proposed in the literature in the context of 2D models [10, 9]. While shuffled hierarchical structures lend themselves better to statistical studies that require averaging across thousands of network realizations, their behavior is mostly identical to those of their deterministic counterparts. For this reason, here we can restrict ourselves to the case of a deterministic hierarchical beam lattice.

For comparison, we introduce randomized non-hierarchical reference patterns as follows. A random beam lattice (RBL) of size $L = 2^n$ is constructed from the same number of $N_{\text{LC}}$ beams and $N_{\text{CL}}$ cross-links as the corresponding DHBL, but the CL beams are distributed randomly over the possible CL sites, leading to an exponential distribution for both the gap linear sizes. Every realization of our RBL consists of a random rearrangement of cross-links.

In order to compute fracture toughness, we also consider systems with pre-existing flaws, in the form of planar crack of length $a$ As depicted in Fig. 2, lattices with pre-existing cracks are modeled through a set of $aL$ missing adjacent LC beams at locations $x_0 \leq x \leq x_0 + a$, $0 \leq y \leq L$, $z = z_0$ where $0 \leq a < L$ is the crack length, $(x_0, 0, z_0)$ denotes the coordinates of the left crack endpoint.

*2.2. Material model*

The constituents of the lattice model are assumed to be straight, identical beams of unit modulus of elasticity, unit length, and square cross-section, which are capable of resisting axial and shear forces and bending moments. There are six degrees of freedom (DOF) at each node, including three translational (node displacements $u$, $v$, and $w$ along the $x$, $y$, and $z$ axes, respectively) and three rotational (rotation angles $\theta_x$, $\theta_y$, and $\theta_z$ about the $x$, $y$, and $z$ axes, respectively), as illustrated in Fig. 3.

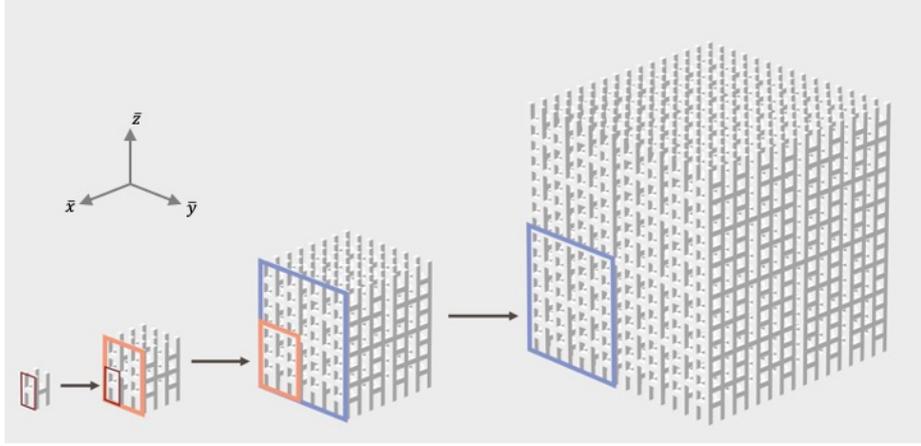

Figure 1: Iterative "bottom-up" construction of a hierarchical lattice. A module of level 1 ("generator") consists of 8 LC beams, plus a CL pane of 4 load-perpendicular (here: horizontal) beams. Level-$n$ systems are constructed by increasing the system size by powers of 2 and replicating the same $8 + 4$ pattern, where the new 8 LC super-beams are now modules which mimic level-$(n-1)$ systems. In figure we show all the cases up to $n = 4$.

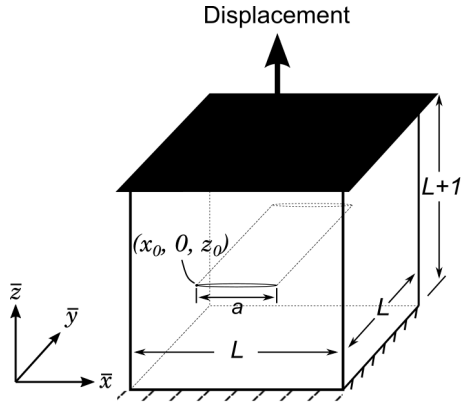

Figure 2: Schematic of simulations on pre-cracked samples. Periodic boundary conditions are applied on free sides of the samples.

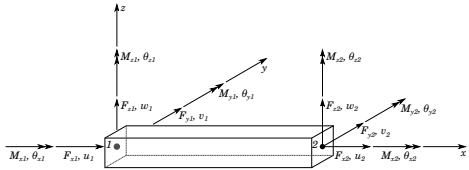

Figure 3: Beam element

The beams are assumed to show a linear elastic mechanical behavior, and Timoshenko beam theory describes their deformation by relating the forces and moments to the associated displacements. Considering a local coordinate system aligned with each beam's axis (see Fig. 3), the displacements and rotations of the beam end nodes are assembled into a local displacement vector $\boldsymbol{u}$, and the forces and moments acting on the end nodes are assembled into a local force vector $\boldsymbol{F}$, resulting in $\boldsymbol{Ku} = \boldsymbol{F}$ for all beams. An explicit expression for a Timoshenko beam's local stiffness matrix $\boldsymbol{K}$ is available in the literature [26].

Assuming quasi-static deformation, we neglect inertial forces. Thus, the global balance equation of the entire lattice takes the form $\bar{\boldsymbol{K}}.\bar{\boldsymbol{u}} = \bar{\boldsymbol{F}}$, where $\bar{\boldsymbol{u}}$ (the global displacement vector) includes all nodal displacements and rotation angles and $\bar{\boldsymbol{F}}$ includes the external forces in the global coordinate system. The global stiffness matrix $\bar{\boldsymbol{K}}$ is formed by first transforming the local stiffness matrices of all lattice beams into the global coordinate system and then assembling them into $\bar{\boldsymbol{K}}$.

*2.3. Failure criterion*

We assume an elastic-brittle mechanical behavior for the beams and remove them irreversibly once a stress-based failure criterion is met. Since the loads on each beam are applied through the end nodes and the beam is in quasi-static equilibrium with small deformations, the magnitude of forces at the beam ends are equal with opposite directions, and the bending moments vary linearly along the beam. Therefore, the beam's stress is maximum at one end, allowing us to check the failure criterion only at the beam end surfaces. We choose a beam-wise failure criterion based on Maximum Distortion Energy Theory (von Mises) as follows:

$$\sigma = \sqrt{\sigma_{xx}^2 + 3\left(\sigma_{xy}^2 + \sigma_{xz}^2\right)} = t\,, \tag{1}$$

where $t$ is the equivalent stress at failure (beam failure threshold). The axial stress $\sigma_{xx}$ in Eq. 1, consisting of tensile/compressive and bending components, at the beam end surface with outward normal $\boldsymbol{n}$ given by

$$\sigma_{xx} = \frac{\boldsymbol{F}.\boldsymbol{n}}{A} + \frac{M_z y}{I_z} + \frac{M_y z}{I_y}\,, \tag{2}$$

where $I_y$ and $I_z$ are the moments of inertia about (local) $y$ and $z$ axes, respectively (see Fig. 3). A force $\boldsymbol{F}$ in the direction of the outward normal of the beam end surface ($\boldsymbol{F}.\boldsymbol{n} = F_x > 0$) yields a positive (tensile) stress contribution and a force in the opposite direction a compressive stress contribution. The shear stresses in Eq. 1, on the other hand, are caused by shear (being maximum at the beam's central axis) and torsional loads (being maximum along the centerline of the beam's outer face) [27]:

$$\begin{aligned}\tau_{\max}(\text{shear}) &= \frac{3V}{2A} \\ \tau_{\max}(\text{torsion}) &= \frac{M_x}{0.208}\,,\end{aligned} \tag{3}$$

where $V$ is the shear force. We choose the bigger absolute $\tau$ value in each transverse direction as the effective shear component:

$$|\sigma_{xy}| = \max(|\frac{3F_y}{2A}|, |\frac{M_x}{0.208}|)$$
$$|\sigma_{xz}| = \max(|\frac{3F_z}{2A}|, |\frac{M_x}{0.208}|) \quad (4)$$

Combining Eq. 1, Eq. 2, and Eq. 4, we have the failure criterion as

$$\sigma/t = 1, \sigma = \sqrt{\left(\frac{F_x}{A} + \frac{M_z y_{\max}}{I_z} + \frac{M_y z_{\max}}{I_y}\right)^2 + 3\left(\max\left(|\frac{3F_y}{2A}|, |\frac{M_x}{0.208}|\right)^2 + \max\left(|\frac{3F_z}{2A}|, |\frac{M_x}{0.208}|\right)^2\right)} \quad (5)$$

which is calculated at both ends of each beam. If the failure criterion is satisfied, the beam is removed irreversibly. A simplified criterion is achieved by excluding the shear force's contribution to failure, which is commonly employed in the literature (see e.g. [28]). This simplification, however, contradicts a Timoshenko model for beam deformation and may grossly underestimate the failure likelihood of lateral connector beams, which mainly transmit shear forces.

*Beam failure thresholds.* To mimic material heterogeneity, the failure thresholds $t$ of beams are randomly assigned using a Weibull probability distribution function with probability density $p(t) = \frac{\beta}{\eta}\left(\frac{t}{\eta}\right)^{\beta-1} \exp\left(-\left(\frac{t}{\eta}\right)^\beta\right)$ with $p(t) \geq 0$, $t \geq 0$, $\beta > 0$, $\eta > 0$, where $\beta$ and $\eta$ are the distribution's shape and scale parameters, respectively. In this study, we choose $\beta = 4$ to represent a moderate degree of disorder and set $\eta = 1/\Gamma(1 + 1/\beta) \approx 1.103263$ so that the failure thresholds $t$ have a mean value of 1. The investigation of the effects of varying $\beta$, in analogy with similar studies in 2D systems [9], is beyond the scope of this work and is being considered for a separate work.

*2.4. Boundary conditions*

The external load is in the form of displacement along LC beams ($\bar{z}$-direction in Fig. 2), imposed through two rigid plates fixed to the top and bottom surfaces of the lattice. Thus, all DOFs of the lattice's bottom and top nodes, except the translational DOF of the top nodes in the loading direction, are fixed. Periodic boundary conditions are imposed on the remaining sides of the lattice (i.e., in $\bar{x}$ and $\bar{y}$ directions).

*2.5. Simulation protocol*

Using a displacement control loading regime, the external displacement is increased until a beam breaks and then kept fixed so that load is re-distributed. The load redistribution (also referred to as stress relaxation) in the absence of the previously broken beam(s) may cause further beam failures without increasing the external load. Once the load redistribution is completed, the external load is increased again, and this process is repeated iteratively until failure is reached, in the form of a system-spanning crack that interrupts load transmission. The displacements of all lattice nodes are calculated at each iteration in the lattice global coordinate system using the global balance equation ($\bar{\boldsymbol{K}}.\bar{\boldsymbol{u}} = \bar{\boldsymbol{F}}$). Then, the global displacements at the nodes are transformed into the beam local coordinate system (see Fig. 3) to calculate the local loads using $\boldsymbol{K}\boldsymbol{u} = \boldsymbol{F}$, and the failure criterion (Eq. 5) is applied as described above.

We note that the pristine/undamaged systems (both DHBL and RBL) consist of $4,860,928$ beams, and the corresponding balance equations are a sparse algebraic system with $\approx 2.9 \times 10^7$ unknowns. In the ramp-up to failure, the evaluation of the balance equations is repeated every time a bream is removed. Every data point in the following is obtained averaging over 10 realizations of the beam network. In each realization, we vary the sequence of thresholds extracted from $p(t)$ as well as (for RBL only) the random positions of CL beams.

## 3. Results

### 3.1. Fracture toughness

To study the effect of hierarchical structure on flaw tolerance, we simulated mode-I-crack propagation in hierarchical (DHBL) and non-hierarchical reference structures (RBL) containing pre-existing cracks of varying length $0 \leq a < L$ (Fig. 2). As depicted in Fig. 4a, hierarchical samples exhibit a super-rough crack surface, which is reminiscent of fracture patterns in bone [3]. The large deflections in crack profiles indicate that the growth of nucleated cracks is arrested and global failure is reached only after a large enough concentration of microcracks coalesces.

Because of this ability to arrest crack growth, hierarchical materials significantly outperform non-hierarchical reference systems in peak stresses and post-peak energy absorption, in situations where pre-existing cracks of length $a$ are considered, as seen on the stress-strain curves in Fig. 4b.

In non-hierarchical system, crack growth is driven by the stress concentrations in the fracture process zone. Failure occurs immediately after the system reaches the peak load $\sigma_p$:

$$\sigma_p = \frac{K_{Ic}}{\sqrt{\pi(a+a_0)}} f\left(\frac{a}{L}\right), \qquad (6)$$

where $K_{Ic}$ is the critical stress intensity factor, $a_0$ is the process zone size [29], and the function $f(a/L)$ accounts for finite size effects. Figure 4 confirms this picture for our RBL (red data points and red lines), with $f(x) = (x/\tan x)^{1/2}$, $x = \pi a/(2L)$ [30]. Here $K_{Ic} \sim \sqrt{G_c E}$ characterizes the material resistance to fracture and depends on the energy release rate $G_c$ (also known as the fracture toughness) and the elastic modulus $E$.

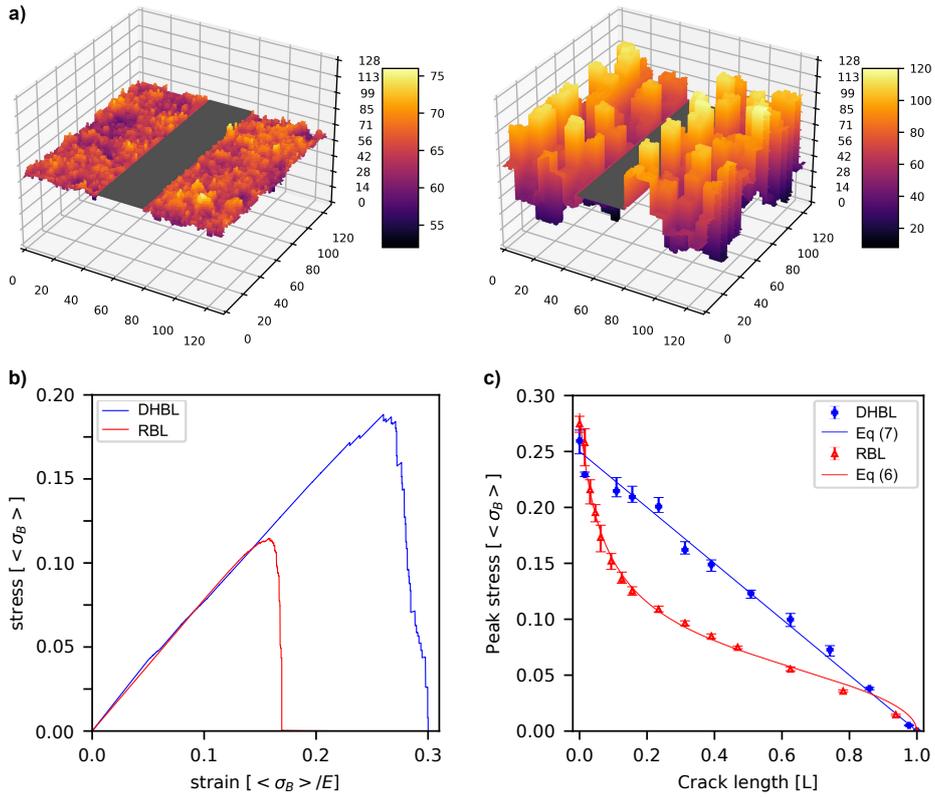

Figure 4: Simulation results for notched hierarchical (DHBL) and non-hierarchical (RBL) lattice variants; (a) fracture surface in RBL (left) and DHBL (right), initial crack of length $a = 30$ is marked in grey, the colorbars indicate height; (b) representative stress-strain curves, crack length $a = 30$, stresses and strains are given as multiples of mean beam failure stress and strain; (c) peak stress as function of crack length, all data are averaged over 10 samples, the error bars indicate the corresponding standard deviation, red and blue lines represent fits according to Eq. 6 and Eq. 7, respectively.

As anticipated, in DHBL fracture proceeds differently. The structural gaps, which in this case are power-law distributed in size, are responsible for arresting crack propagation and altering long range stress redistributions. As a consequence, DHBL display a reduced sensitivity to pre-existing cracks, with peak loads that decrease only linearly with the notch size $a$ as

$$\sigma_p = \sigma_0 \left(1 - \frac{a}{L}\right), \tag{7}$$

where $\sigma_0$ is the peak stress in the absence of pre-existing cracks. In passing, we note that the case of $a \approx 0$ is the only exception, where HBL fall short of non-hierarchical systems: the relevant aspect of biomimetic hierarchical structures is their ability to cope with pre-existing flaws, rather than their peak performance under ideal conditions.

*3.2. Characteristics of fracture surfaces*

*Absence of crack growth.* To confirm our statement that crack growth is arrested in DHBL, we started by looking at damage nucleation and growth patterns of un-notched sample as depicted in Fig. 5 for a typical DHBL (left) and RBL (right) Fracture in non-hierarchical materials is caused by the nucleation and propagation of a crack that becomes critical at the system's peak load and spreads throughout the system due to crack-tip stress concentrations, resulting in an abrupt and catastrophic failure. In DHBL, on the other hand, LC modules fail individually and their state does not propagate to neighboring modules: the vertical deflections indicate that a growing crack has reached a gap and has stopped. The hierarchically distributed gaps inhibit crack propagation at all scales, resulting in widely separated flaws coalescing into a super-rough fracture profile.

*Crack roughness.* The statistical analysis of crack surface patterns allows us to further understand the machanisms of failure onset and propagation. We consider the fracture surfaces as a function $\bar{z}(\bar{x}, \bar{y})$ in the global coordinate system. We start by computing the crack roughness in terms of the scale-dependent standard deviation

$$\sigma_l = \left\langle \langle (\bar{z}(\bar{x}, \bar{y}) - \langle \bar{z}(\bar{x}, \bar{y}) \rangle_l)^2 \rangle_l^{1/2} \right\rangle_{L,N} \tag{8}$$

in both global $\bar{x}$ and $\bar{y}$ directions, where $\langle \ldots \rangle_l$ denotes the average over a window of length $l$, and $\langle \ldots \rangle_{L,N}$ is the average over all windows contained in the sample cross section of length $L$, as well as over all samples in an ensemble of $N$ simulations with different realizations of a given microstructure.

We investigate the possibility of crack profiles being self affine. In this context, self affinity refers to invariance under the scaling transformation $\bar{x} \to \lambda \bar{x}, \bar{y} \to \lambda \bar{y}, \bar{z} \to \lambda^H \bar{z}$, where $H$ is the Hurst exponent [9]. Self affinity, in particular via a non-trivial value of $H$ that cannot be inferred from the structure, indicates that the fracture process is affected by emergent features and correlations of the dynamics at hands. Evaluating $\sigma_l$ allows us to verify if these scaling relationships hold, and to evaluate $H$ in case they do.

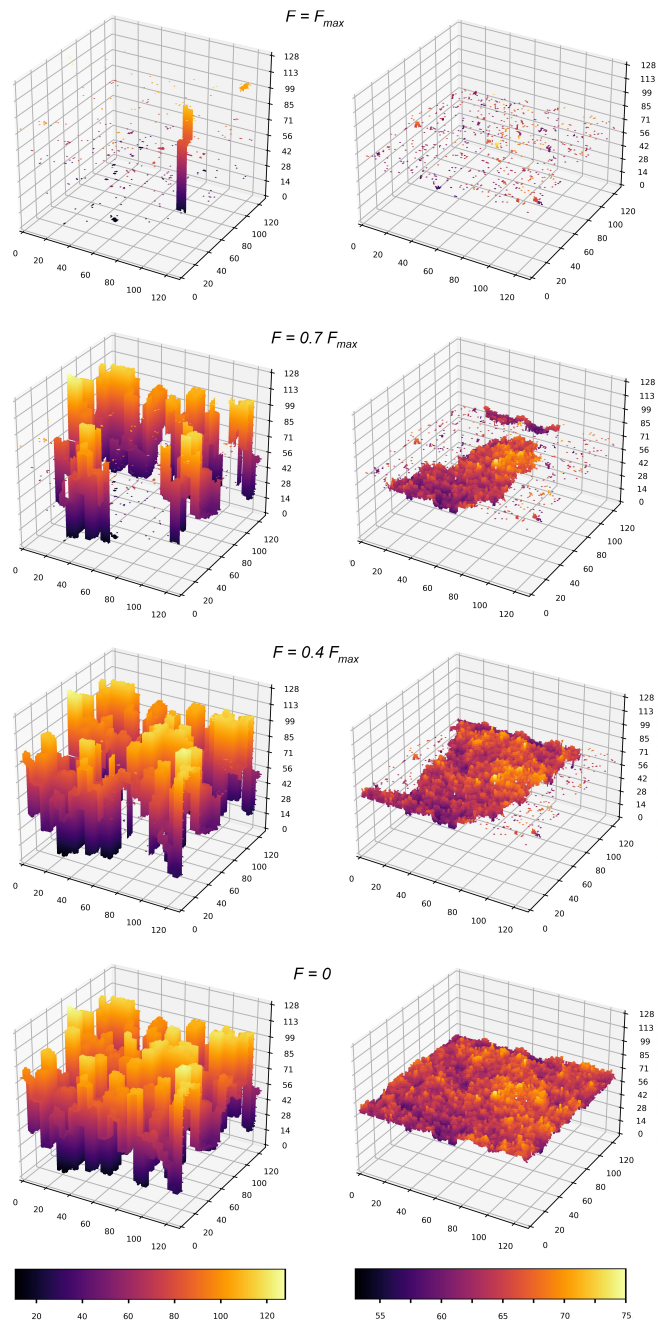

Figure 5: Typical damage growth patterns, from the system's peak load (top) to global failure (bottom); left column: DHBL, right column: RBL.

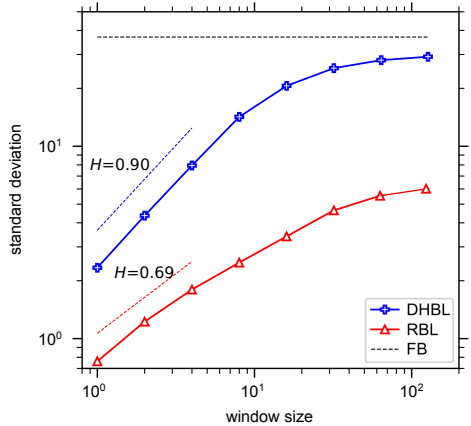

Figure 6: Standard deviation $\sigma_l$ vs averaging length $l$ for different 3D lattice variants.

Fig. 6 illustrates the values of $\sigma_l$ for DHBL and RBL. RBL show a self-affine scaling with a non trivial Hurst exponent $H = 0.69$, confirming that in heterogenous non-hierarchical systems growth is controlled by dynamic correlations [31, 20]. On the other hand, in DHBL we record an apparent $H = 0.90$. This result can be interpreted as follows: an exponent of 1 derives from the fact that hierarchical lattices with an infinite number of hierarchical levels are invariant under the transformation $\bar{x} \to 2\bar{x}$, $\bar{y} \to 2\bar{y}$, $\bar{z} \to 2\bar{z}$, and the same must be valid for the associated crack profiles. Contrarily to the case of RBL, $H$ is compatible with a prediction based exclusively on structural considerations. The apparent exponent in our simulations is slightly lower ($H = 0.9$) due to the finite-size and boundary effects, and as a consequence the scaling behavior is limited to small $l$, and abruptly interrupted for larger $l$.

Finally, for comparison, Fig. 6 also shows the value $\sigma_l = L/\sqrt{12}$ for an equal load sharing fiber bundle model with large fibers of length $L$, obtained under the assumption that $\bar{z}(\bar{x}, \bar{y})$ values are random variables in the interval $[0, L]$.

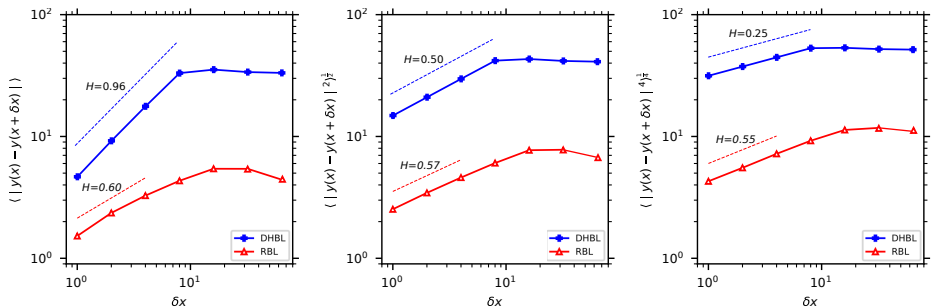

Figure 7: Structure factors $C^m$ for different values of $m$ and different 3D lattice variants. Left column shows $C^1$, middle column $C^2$, right column $C^4$.

In order to further explore the origin of $H$ in DHBL, we perform a multi-scaling analysis. Denoting $\Delta(\bar{x}, \bar{y}, l)$ as the height difference $|z(x,y)-z(x+l,y)|$ and $|z(x,y)-z(x,y+l)|$ between the points on a profile separated by a distance $l$, the so-called structure factors of order $m$ are then computed from

$$C^m(l) = \langle [\Delta(\bar{x}, \bar{y}, l)]^m \rangle_{L,N}^{1/m}. \tag{9}$$

A self-affine crack surface of Hurst exponent $H$ would exhibit $C^m(l) \propto l^H$ for all $m$. Fig. 7 confirms that indeed that this condition is approximately met for crack profiles in non-hierarchical lattices RBL, based on the results of a multi-scaling analysis. In hierarchical lattices (DHBL), however, the scaling is not self-affine since the apparent Hurst exponent for different structural factors $C^m$ is dependent on the exponent $m$ with $H_m = 1/m$.

We can finally rule out the self-affinity hypothesis for DHBL. At least under the current assumptions of local strength fluctuations (encoded in the choice of $\beta = 4$), we observe no explicit evidence of emergent behavior and the remarkable super-rough fracture surfaces encountered in these systems seem to be a direct consequence of the hierarchical structure and its fractal-like organization.

## 4. Conclusions

In this paper, we presented the first simulations of a large-scale, three-dimensional model for semi-brittle fracture in biomimetic hierarchical materials, using full-tensorial constitutive laws and failure criteria. The degree of accuracy in our modeling approach allows us to draw conclusions about the fracture behavior of hierarchical materials, which, in the past, could only be explored under significant simplifications.

Our results confirm the long standing view, originated in the context of bone fracture, according to which hierarchical microstructures arrest crack growth. Starting from this observation, we provide compelling numerical evidence of enhanced fault tolerance, with respect to reference non-hierarchical systems. Our DHBL models outperform comparable random lattices, displaying higher peak loads when dealing with previosuly accumulated damage. A statistical analysis of the fracture surfaces allows us to conclude that, under our assumptions, this remarkable behavior shows little dependence on dynamic correlations, and is for the most part controlled by the network structure.

The role played by the hierarchical microstructure, arresting crack growth and thus enhancing fault tolerance, represents a significant advantage in biomaterial modeling, as it paves the way to techniques of performance tuning based on microstructure design and synthesis. In a realistic scenario of this type, our DHBL model allows the parametrization of an additively manufactured hierarchical network of brittle beams or struts, by acting on i) the Weibull distribution exponents $\beta$ to control the spread in local strength fluctuations, and ii) the actual constitutive laws and failure criteria. We note that in the present work we focused on the specific value of $\beta = 4$, and on a specific choice of constitutive/failure behavior. An extensive study of the role of these additional variables is left for future work. The role of fluctuations in local strength, in particular, deserves special attention. Based on results on two dimensional systems [9], we can expect that in the case of much larger fluctuations ($\beta \approx 1$), disorder and stochastic fluctuations might become as prominent the hierarchical structure in shaping the fracture process.